\pdfoutput=1

\documentclass[conference]{IEEEtran}

\ifCLASSINFOpdf
\else
\fi
\usepackage[T1]{fontenc}
\usepackage{newtxtext,newtxmath}
\usepackage{microtype}
\usepackage[scaled=.84]{beramono}

\usepackage{relsize}
\usepackage{xspace}
\usepackage[normalem]{ulem}

\usepackage{flushend}

\usepackage[nounderscore]{syntax}

\usepackage{graphicx}
\usepackage{multicol}
\usepackage{multirow}
\usepackage[]{algorithm2e}
\usepackage{fixltx2e}
\usepackage{algorithmic}
\usepackage{listings}
\usepackage{booktabs}
\usepackage{verbatim}
\usepackage{paralist}
\usepackage{subcaption}
\usepackage{tabulary}

\usepackage[
  bookmarksnumbered=true,
  pdfstartview=FitV,
  bookmarksopen,
  bookmarksopenlevel=2]
{hyperref}
\usepackage[all]{hypcap}

\usepackage{cleveref}

\usepackage{wrapfig}

\usepackage[xspace]{ellipsis}

\lstdefinelanguage{C++1y}{
  alsolanguage=C++,
  escapechar=@,
  breakatwhitespace=true,
  morekeywords = {
    alignof, decltype, concept, axiom, requires, property
  }
}

\lstnewenvironment{programnb}[1][\small]
{
  \lstset{
    style=C++,
    basicstyle=#1\sffamily,
    keywordstyle=#1\sffamily,
    commentstyle=#1\sffamily,
  }
}
{ }

\newtheorem{mydef}{Definition}
\newtheorem{mynot}{Notation}

\newtheorem{myproposition}{Proposition}

\usepackage{tikz}

\definecolor{lightblue}{rgb}{0,0,0}
\lstdefinelanguage{C++custom}
{
escapeinside={/*@}{@*/},breaklines=true,breakatwhitespace=true,%
basicstyle=\color{lightblue},keywordstyle=,%
lineskip=-.05\baselineskip,morekeywords={pattern,in,is,to,from,out_edges,adj,once,fixed_point,vertex_property,edge_property,vertex,edge,meta,auto,concept,requires,concept_map},%
alsolanguage=C++,
literate = %
  {\{}{\smaller{$\{$}}{1}%
  {\}}{\smaller{$\}$}}{1}%
  {<=}{$\leq$}{1}%
  {-}{--}{1}%
}

\lstdefinestyle{numbers}
{xleftmargin=8pt,numbers=left, numberstyle=\tiny,%
stepnumber=1, numbersep=4pt}

\lstdefinestyle{frametb}
{frame=tb}

\lstdefinestyle{bold-keywords}{keywordstyle=\bfseries}

\lstnewenvironment{myverb}[1][\small]{\lstset{
    columns=fullflexible,
    basicstyle=\color{lightblue}#1\ttfamily,%
    breaklines=true,
  }}{}

\lstnewenvironment{cplus}[1][\footnotesize]{\lstset{language=C++custom,%
    style=numbers,basicstyle=\color{lightblue}#1\ttfamily,%
    keywordstyle=#1\ttfamily,%
    style=bold-keywords,style=frametb}}{}

\lstnewenvironment{cplusuf}[1][\small]{\lstset{language=C++custom,%
    style=numbers,basicstyle=\color{lightblue}#1\ttfamily,%
    keywordstyle=#1\ttfamily,%
    style=bold-keywords,frame=t
  }}{}

\lstnewenvironment{mjava}[1][\small]{\lstset{language=Java,%
    escapeinside={/*@}{@*/},style=numbers,basicstyle=\color{lightblue}#1\ttfamily,%
    keywordstyle=#1\ttfamily,%
    style=bold-keywords,style=frametb}}{}

\lstnewenvironment{cplusnln}[1][\small]{\lstset{language=C++custom,
    xleftmargin=8pt,xrightmargin=8pt,basicstyle=\color{lightblue}#1\ttfamily,%
    keywordstyle=#1\ttfamily,%
    style=bold-keywords,style=frametb}}{}

\makeatletter
\lstnewenvironment{code}[1][\footnotesize]{\lstset{language=C++custom,columns=fullflexible,%
    xleftmargin=8pt,xrightmargin=8pt,basicstyle=\color{lightblue}#1,%
    keywordstyle=#1,style=numbers,%
    style=bold-keywords,mathescape=true}}{\@endparenv}
\makeatother

\providecommand{\codeinl}[2][\small] 
{{\lstinline[language=C++custom,breaklines=false,columns=fullflexible,%
basicstyle=\color{lightblue}#1\sffamily
,mathescape=true,keywordstyle=#1\sffamily]@#2@}}%

\providecommand{\cplusinl}[2][\normalsize] 
{{\lstinline[language=C++custom,breaklines=false,columns=fullflexible,%
basicstyle=\color{lightblue}#1\ttfamily
,keywordstyle=#1\ttfamily]@#2@}}%

\lstdefinestyle{cppmarkers}{rangeprefix=/*\#\ ,%
includerangemarker=false,%
rangesuffix=\ \#*/}%

\lstdefinelanguage{Haskell-custom}
{
escapeinside={--@}{@--},breaklines=true,breakatwhitespace=true%
language=Haskell,basicstyle=\color{lightblue}\ttfamily,keywordstyle=\ttfamily,%
morekeywords={class,instance,type,newtype,data,where,deriving,import},%
lineskip=-.1\baselineskip,morekeywords={concept,requires,concept_map},
literate={+}{{$+$}}1 {/}{{$/$}}1 {*}{{$*$}}1 {=}{{$=$}}1
               {>}{{$>$}}1 {<}{{$<$}}1 {\\}{{$\lambda$}}1
               {\\\\}{{\char`\\\char`\\}}1
               {->}{{$\rightarrow$}}2 {>=}{{$\geq$}}2 {<-}{{$\leftarrow$}}2
               {=>}{{$\Rightarrow$}}2
               {\ .}{{$\circ$}}2 {\ .\ }{{$\circ$}}2
               {>>}{{>>}}2 {>>=}{{>>=}}2
               {|}{{$\mid$}}1
             }

\lstnewenvironment{hask}[1][\small]{\lstset{language=Haskell-custom,%
    style=numbers,basicstyle=\color{lightblue}#1\ttfamily,keywordstyle=#1\ttfamily,%
    style=bold-keywords,style=frametb}}{}

\lstdefinestyle{markers}{rangeprefix=\{-\:\ ,%
includerangemarker=false,%
rangesuffix=\ \:-\}}%

\lstdefinestyle{C++}
{
  language=C++1y,
  columns=fullflexible,
  breaklines=true,
}

\newcommand\Dstepping{$\Delta$-stepping\xspace}
\newcommand\DStepping{$\Delta$-Stepping\xspace}

\usepackage{mathtools}

\lstdefinelanguage{CASL}{
  escapechar=@,
  breakatwhitespace=true,
  morekeywords = {
    spec, sort, then, op, ops, var, vars, pred, end
  }
}

\lstnewenvironment{casl}[1][\small]
{
  \lstset{
    language=CASL,
    basicstyle=#1\sffamily,
    keywordstyle=#1\sffamily\bfseries,
    commentstyle=#1\sffamily,
    columns=fullflexible,
    breaklines=true,
  }
}
{ }

\usepackage{pdfcomment}
\usepackage[textwidth=1.3\marginparwidth]{todonotes}

\defineavatar{MZ}{author=MZ,color=yellow}
\defineavatar{AJ}{author=AJ,color=blue}
\defineavatar{RO}{author=RO,color=green}
\defineavatar{MRT}{author=MRT,color=red}

\providecommand{\RO}[2][]{}
\providecommand{\AJ}[2][]{}

\newcommand{\mli}[1]{\ensuremath{\mathit{#1}}}
\providecommand{\WorkItemSet}{\ensuremath{\mli{WorkItem}}\xspace}
\providecommand{\WorkItem}{\ensuremath{\mli{workitem}}\xspace}
\providecommand{\WorkItems}{work items\xspace}
\providecommand{\wis}{\ensuremath{\mli{wis}}\xspace}
\providecommand{\WIS}{\ensuremath{\mli{WIS}}\xspace}
\providecommand{\Condition}{\ensuremath{\mli{C}}\xspace}
\providecommand{\StateSet}{\ensuremath{\mli{Q}}\xspace}
\providecommand{\StateUpdate}{\ensuremath{\mli{U}}\xspace}
\providecommand{\Constructor}{\ensuremath{\mli{N}}\xspace}

\providecommand{\PF}{\textit{$\pi$}\xspace}
\providecommand{\citep}{ \cite}
\providecommand{\algoref}{ Algorithm~\ref}
\providecommand{\defref}{ Definition~\ref}
\providecommand{\secref}{ Section~\ref}
\providecommand{\propref}{ Proposition~\ref}
\providecommand{\figref}{Figure~\ref}
\newcommand{\powerset}[1]{\mathcal{P}(#1)}

\newcommand\eg{e.g.,\xspace}

\usepackage{float}

\floatstyle{ruled}
\newfloat{algorithm}{tbh}{lop}
\floatname{algorithm}{Algorithm}

\usepackage{caption}
\usepackage[subrefformat=parens,labelformat=parens]{subcaption}

\providecommand{\linesref}[2]{\hyperref[#1]{Lines~\ref*{#1}--\ref*{#2}}}
\providecommand{\linesandref}[2]{\hyperref[#1]{Lines~\ref*{#1}} \hyperref[#2]{and~\ref*{#2}}}
\providecommand{\lineref}[1]{\hyperref[#1]{Line~\ref*{#1}}}

\begin{document}
%
\title{Families of Distributed Memory Parallel Graph Algorithms from Self-Stabilizing Kernels--An SSSP Case Study}


\author{\IEEEauthorblockN{Thejaka Kanewala\IEEEauthorrefmark{1}\IEEEauthorrefmark{2},
Marcin Zalewski\IEEEauthorrefmark{1}, Martina Barnas\IEEEauthorrefmark{2} and
Andrew Lumsdaine\IEEEauthorrefmark{1}}
\IEEEauthorblockA{\IEEEauthorrefmark{1}Pacific Northwest National Laboratory \& University of Washington,
Seattle, WA, USA.\\
\IEEEauthorrefmark{2}School of Informatics \& Computing,
Indiana University\\
Bloomington, IN, USA.\\
Email: \IEEEauthorrefmark{1}\{thejaka.kanewala, marcin.zalewski, andrew.lumsdaine\}@pnnl.gov,
\IEEEauthorrefmark{2}\{thejkane, mbarnas\}@indiana.edu}}


\maketitle

\begin{abstract}

  Self-stabilizing algorithms are an important because
  of their robustness and guaranteed convergence.
  Starting from any
  arbitrary state, a self-stabilizing algorithm is guaranteed to converge to a legitimate
  state.
  Those algorithms are not directly amenable to solving distributed graph
  processing problems when performance and scalability are important.
%
  In this paper, we show the ``Abstract Graph Machine'' (AGM) model that can be used
  to convert self-stabilizing algorithms into forms suitable for distributed
  graph processing.
  An AGM is a mathematical
  model of parallel computation on graphs that
  adds work dependency and ordering to self-stabilizing algorithms.
  Using the AGM model we show that some of the existing distributed
  Single Source Shortest Path (SSSP) algorithms
  are actually specializations of self-stabilizing SSSP.
  We extend the AGM model to apply more fine-grained orderings at different spatial
  levels to derive additional scalable variants of SSSP algorithms, essentially
  enabling the algorithm to be generated for a specific target architecture.
  Experimental results show that this approach can generate new algorithmic variants
  that out-perform standard distributed algorithms for SSSP.

\end{abstract}


%
\IEEEpeerreviewmaketitle

\section{Introduction}
\label{sec:intro}
Most of the existing parallel algorithms are developed based on
Parallel Random Access Machine (PRAM)~\cite{fortune1978parallelism}
model.
While PRAM is a simple machine model it does not consider
factors that are significant in distributed memory system
(\eg overhead of synchronization, remote message communication
etc.). Further, algorithms designed for PRAM may assume global
data structures and subgraph computations are efficient,
hence their cost is not counted into algorithm performance.
While these algorithms perform well in shared memory systems
they tend to perform poorly on distributed memory systems.

\textit{Self-stabilizing} graph algorithms rely on local information
to solve graph problems.
In self-stabilizing algorithms
every vertex in the graph is associated with a state.
Whenever, there is a state change,
neighboring
vertices are notified via edges.
Self-stabilizing algorithms does not use
global data structures and does not rely
on operations such as subgraph computations (these
operations are expensive in a distributed environment).
The fact that, self-stabilizing
algorithms rely only on local information
and does not assume global data structures
motivate
us to investigate the applicability of self-stabilizing
graph algorithms for large scale static graph
processing.

A self-stabilizing algorithm consists of set of
rules. Every rule has a condition. A rule
is evaluated only if its condition evaluates to
true.
A self-stabilizing algorithm reaches a legitimate
state irrespective of its initial state.

Self-stabilizing graph
algorithms have been introduced for number of graph
applications including \emph{Single Source Shortest Path},
\emph{Breadth First Search} (BFS),
\emph{Spanning Tree Construction},
\emph{Maximal Independent Set} and
\emph{Graph Coloring}. A survey of self-stabilizing graph algorithms
is presented in\citep{guellati2010survey}.

The strategy for updating vertex states is defined by
an \emph{execution model}. In self-stabilization, those
execution models are called \emph{demons}. We find three types
of demons in self-stabilizing algorithms: \emph{central demon},
\emph{synchronous demon} and \emph{distributed demon}.
In a central demon algorithm, only one vertex can update the state
at a time. While synchronous demon updates all the vertex
states at the same time, a distributed demon select a
subset of vertices to update states at the same time.
Since our main focus is to reduce global synchronization
and to rely on ``local'' data for processing, we only
consider distributed demon algorithms.



A distributed demon self-stabilizing
Single Source Shortest Path algorithm
is presented in\algoref{SSSP_relax} (discussed in~\citep{huang2002self}).
At stabilization this algorithm
will have the minimum distance to every vertex, from
a given source vertex.
Self-stabilizing algorithms describe
the algorithm using
rules.
A rule consists of a condition and an action, when
condition evaluated true action is invoked.
Rules in the\algoref{SSSP_relax}
have the format;
\emph{current state} $\longrightarrow$ \emph{new state}.

\begin{algorithm}
\algsetup{linenosize=\scriptsize}
\caption{Self-stabilizing SSSP for distributed demon}
\label{SSSP_relax}
\begin{algorithmic}[1]
\STATE \{For the source r \}
\STATE R0: $d(r) \neq 0 \longrightarrow d(r):=0$
\STATE \{For node $i \neq r$ \}
\STATE R1: $d(i) \neq min_{j \in N(i)}(d(j)+w(i,j)) \longrightarrow d(i):=min_{j \in N(i)}(d(j)+w(i,j))$
\end{algorithmic}
\end{algorithm}

Note that in the algorithm, $d(i)$ represents the state of
vertex $i$ and $N(i)=\{i \in V | (i,j) \in E\}$ stands for
set of all neighbours of vertex $i$. The pre assigned weight
for an edge is denoted by $w(i,j)$.

The algorithm consists of two rules. The first rule is only
applicable to the source. It says, if the source ``distance
state'' is not 0, then the ``distance state'' should be transferred
to 0. The second rule is activated only if the current distance
of the vertex is not
equal to the minimum of its neighbours distance plus the
weight of the edge connecting the neighbour. So, the legitimate
state of the system is $d(r)=0$ ($r$ is the given source)
and $\forall i \neq r, d(i) = min_{j \in N(i)}(d(j)+w(i,j))$.~\citep{huang2002self} proves
that this algorithm ultimately stabilizes and $d(i)$
represents the shortest distance from the source, for each vertex $i$ at stabilization.

Another important aspect of self-stabilizing algorithms
is the \emph{atomicity} requirement.\algoref{SSSP_relax}
requires to querying vertex $i$'s state and its
neighbors state and updating vertex $i$'s state (if the rule evaluates to
true for vertex $i$) in a single atomic step.
In self-stabilization, the requirement to query the current vertex
state and its neighbours states and update the current vertex
state in a single atomic step is called \emph{composite atomicity}.
The \emph{Read-Write} atomicity is another form of atomicity
discussed in detail in\citep{dolev1993self}.
In this paper we consider self-stabilizing algorithms
with \emph{composite atomicity}.


Using self-stabilizing algorithms for
static graph analysis is challenging
due to several reasons: 1. Self-stabilizing
algorithms are designed for streaming contexts in which
those algorithms do not terminate after stabilizing;
2. To implement the composite atomicity requirement we need to
lock the current vertex as well as its neighbors. This will generate a
lot of synchronization regions and involves a significant number of
locking/unlocking operations, hence it reduces performance especially
in distributed execution;
3. A naive implementation of the self-stabilizing SSSP would require
program to iterate through all the vertices and apply rules R0 and R1
until they reach the legitimate state. When processing a large graph,
such an implementation would generate a massive amount of unnecessary work
and would perform very poorly.

To overcome the challenges discussed above, we model
self-stabilizing SSSP\algoref{SSSP_relax} using an \emph{Abstract Graph Machine} (AGM)\citep{2016agm}.
An AGM is a model for
designing distributed memory parallel
graph algorithms.
An AGM essentially converts the self-stabilizing algorithm to a
\emph{stabilizing} algorithm by adding work dependency 
and uses ordering to reduce the amount of work. A stabilizing algorithm starts from a
specific initial state where as self-stabilizing algorithm can execute the algorithm from any
initial state.
The modeled stabilizing algorithm is in a suitable form for distributed graph
processing.
We also
show that some of the existing well-known distributed SSSP
algorithms are specific versions of the modeled self-stabilizing algorithm
in~\ref{SSSP_relax}.

We further enhance the AGM model to incorporate architecture dependent
spatial characteristics to generate less synchronized orderings.
The extended model is called \emph{Extended AGM} (EAGM). Using
EAGM we generate \emph{nine} variations of SSSP algorithms
and compare their performance to standard distributed SSSP
algorithms under three different types of graph inputs.
Our results show that some of the generated algorithm variations
perform better compared to standard distributed SSSP algorithms.






\section{Self-Stabilizing SSSP \& AGM}
\label{ssandsssp}




In the following we discuss several
important aspects when modeling the self-stabilizing
SSSP algorithm using AGM. More specifically, our
discussion is focused on termination,
composite atomicity and ordering.

Self-stabilizing algorithms are not
iterative algorithms neither the AGM algorithms.
While self-stabilizing algorithms does not
terminate they go to an idle state when
states are stabilized but AGM algorithms
terminate at stabilization.
Algorithm termination depends on the amount
of active work available in the system.
Our implementations use standard termination
detection algorithms to count active
work available in the system alive. As long as
there are state changes there will be active work
available. When active work is zero there are
no more state changes and guarantees that
the algorithm state is stabilized.

Under composite atomicity \algoref{SSSP_relax}
needs to query states from neighbors,
query the current state and update current
the vertex state in a single atomic step.
This requires synchronization between
neighboring vertices.
However, for the SSSP algorithm in~\ref{SSSP_relax}, the synchronization requirement
can be alleviated due to the \emph{monotonic}
 function ``min''. In rule 1,
neighbors states are processed in the
``min'' function. The ``min'' function selects the minimum
distance + weight value for all neighbors, irrespective
of the order states pushed in. Therefore we only need to
maintain the state of the current vertex.

The AGM model makes state transitions
based on work events. We define a unit of
work to be a pair, $<v, s_v>$ where $v \in V$
and $s_v$ is the state associated with $v$. We
call a unit of work a \WorkItem and we denote
all the possible \WorkItems generated by an algorithm
as the set \WorkItemSet.
In the AGM model, every-time a state
is updated new \WorkItems are generated.

Before processing, new \WorkItems
are ordered. In an AGM
ordering is defined as a
\textit{strict weak ordering}.
The strict weak ordering
divide \WorkItems into different
equivalence classes.

\section{Abstract Graph Machine (AGM)}
An \emph{Abstract Graph Machine}(AGM) consists of a
definition of a \emph{\WorkItemSet},
an \emph{initial \WorkItem set},
a set of \emph{states},
a \emph{processing function}
and a \emph{strict weak ordering} relation.
The processing function takes a \WorkItem ($\in \WorkItemSet$)
as an input and may produce
0 or more \WorkItems. Further, the processing function
may change values associated with states. The strict weak ordering
relation order \WorkItems into a set of \emph{ordered}
(induced) \emph{equivalence
classes}.

The AGM model denotes a \WorkItem ($\in \WorkItemSet$) as a \emph{tuple}.
A tuple's first element stores a vertex and the rest of the positions
store the state/s and ordering attribute values.
For example, the self-stabilizing
SSSP algorithm stores a vertex and a distance in a \WorkItem.
The \WorkItem values are populated by the processing function.
The size of the tuple (i.e. the number of additional elements)
is determined by the states in the algorithm and the ordering attributes
used in the AGM formulation.

To access tuple elements we use the \emph{bracket operator};
e.g., if $w \in \WorkItemSet$ and
if $w$ = \textless$v, p_0, p_1\dots, p_n$\textgreater
then $w[0]$ = v and $w[1]$ = $p_0$ and $w[2]$ = $p_1$,
etc. The \WorkItem data (i.e. tuple elements) are read
by the processing function. After reading values,
the processing function can change the states associated
with the vertex in the \WorkItem.

An AGM maintains state values as \emph{mappings}(functions).
The domain of the state mappings is the set $V$. The co-domain
depends on the possible values that can be held in states.
For example, in the self-stabilizing SSSP algorithm
the state mapping is $distance : V \longrightarrow \mathbb{R}$.
In AGM terminology,
accessing a state value associated with a vertex (or edge) ``v''
is denoted as ``mapping\_name[v]'' (e.g., distance[v]).

In addition to state mappings, algorithms also
use \emph{read-only} properties. These properties
usually hold graph data and are interpreted as functions,
e.g., edge weights ($weights : E \longrightarrow \mathbb{R}$).
In the abstraction we treat those read-only properties
as part of the graph definition.
In terms of the syntax, to read values,
the AGM model does not distinguish
between
read-only properties and state mappings.

Both state
mappings and properties are accessed within a processing
function and the processing function only reads from
the properties while processing function may modify
state mappings when processing a \WorkItem.
The processing function updates
the state value
associated with a vertex
in a \emph{single atomic step}.

The logic inside the processing function
is analogous to the code
that runs infinitely on every process in a
self-stabilizing algorithm. However, in the AGM model a single
process may run multiple instances of processing
functions (e.g., distributed shared memory model). The logic inside
the processing function is based on the rules
defined for self-stabilizing algorithms but is
modified to work with \WorkItems.
A processing function (\PF) takes a \WorkItem
and may produce more \WorkItems based on the logic defined inside
the \PF. In our formalism we treat states as \textit{side effects},
in the sense they are not passed as explicit
arguments but subject to change when executing \PF.
We will denote set of states using \StateSet and we will
use $\powerset{W}$ to denote the powerset of set $W$.
Then, mathematically the \PF is declared as
$\PF : \WorkItemSet \longrightarrow \powerset{\WorkItemSet}$.

The processing function defines the
basic logic of an algorithm.
It consists of a set of \emph{statements} ($St$).
A statement specifies
a \textit{condition} based on
input \WorkItem and/or states
($\mli{\Condition} : \WorkItemSet \longrightarrow \{True,False\}$) and
an \textit{update to states}
($\mli{\StateUpdate} : \WorkItemSet \longrightarrow \{True, False\}$) and
how a \textit{output \WorkItems} ($w_{new}$)
should be constructed
($\mli{\Constructor} : \WorkItemSet \longrightarrow \powerset{\WorkItemSet}$).
The \Constructor of a statement is evaluated
only if its \Condition (condition) and \StateUpdate
are evaluates to $True$. We distinguish between $U$ and $C$,
since $U$ may have side effects where it updates states but
$C$ does not create any side effects.

An abstract version of \PF is formally defined in\defref{pf}.


\begin{mydef}\label{pf}
  $\mli{\PF} : \WorkItemSet \longrightarrow \powerset{\WorkItemSet}$
  \[ \mli{\PF}(w) =  \bigcup_{s_i \in St} s_{i}(w)
  \] \\
  \text{where; }$\mli{s_i} : \WorkItemSet \longrightarrow \powerset{\WorkItemSet}$
  \[ \mli{s_i}(w) =
  \begin{cases}
  \{ w_{new} | w_{new} \in \Constructor(w)
  \\
  \; \; \text{ if } \text{\Condition(w) is True \& $\StateUpdate(w)$ is True} \}
  \\
  \{\} else
  \end{cases}
  \]
\end{mydef}

\begin{figure}
\centering
\includegraphics[width=.95\linewidth]{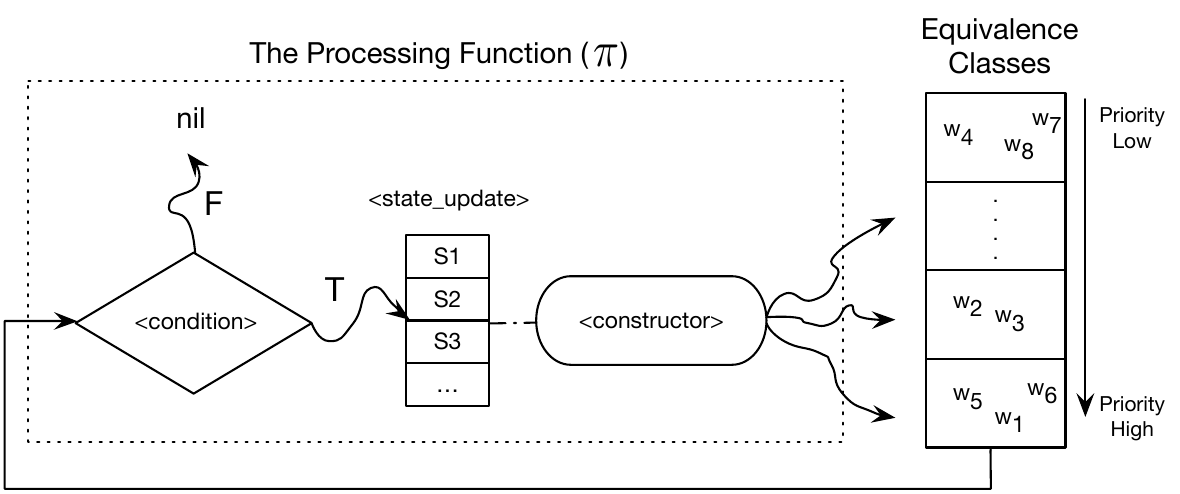}
\caption{An overview of the Abstract Graph Machine}
\vspace{-1ex}
\label{fig:agm}
\end{figure}

The output \WorkItems of
a processing function are ordered according to the strict weak ordering
defined on \WorkItemSet. The ordered \WorkItems are again
fed into the processing function in the order they appear.
The interaction between
the processing function and
the ordering is graphically depicted in the Figure~\ref{fig:agm}.

The strict weak ordering relation (denoted by $<_{\wis}$)
must satisfy the following properties; \AJ{Or constraints ?}
\begin{enumerate}
\item For all $w \in \WorkItemSet$, $w \nless_{\wis} w$.
\item For all $w_1, w_2 \in \WorkItemSet$, if $w_1 <_{\wis} w_2$, then $w_2 \nless_{\wis} w_1$.
\item For all $w_1, w_2, w_3 \in \WorkItemSet$, if $w_1 <_{\wis} w_2$ and $w_2 <_{\wis} w_3$, then $w_1 <_{\wis} w_3$.
\item For all $w_1, w_2, w_3 \in \WorkItemSet$, if $w_1$ not comparable with $w_2$ and $w_2$ not comparable with $w_3$, then $w_1$ is not comparable with $w_3$.
\end{enumerate}

Properties 1 and 2 states that the strict
weak ordering relation is \textbf{not} \emph{reflexive} and is
\emph{anti-symmetric}. Property 3 denotes the
\emph{transitivity} of the ``comparable \WorkItems''
and Property 4 states that transitivity is preserved among
non-comparable elements in the \WorkItem set.
These properties give rise to an \emph{equivalence}
(i.e. non-comparable \WorkItems belong to the same
equivalence class)
relation defined on \WorkItemSet, hence the strict weak ordering
relation partitions
the complete \WorkItemSet. Since \WorkItems in
different equivalence classes are comparable, the
strict weak ordering relation defined on the set \WorkItemSet
\emph{induces an ordering} on generated equivalence
classes. In general, there are several ways to define the induced
ordering relation (denoted $<_{\WIS}$). For our work we use the definition
given in the\defref{def-graph-aws-relation}.

\begin{mydef}\label{def-graph-aws-relation}
  $<_{\WIS}$ is a binary relation defined on $\powerset{\WorkItemSet}$, such that if $W_1, W_2 \in
  \powerset{\WorkItemSet}$ then; $W_1 \le_{\WIS} W_2
  \; iff \\
  \; forall \; w_1 \in W_1 \; and \; forall \; w_2 \in W_2 ; w_1 <_{\wis} w_2$.
\end{mydef}



Having defined all supporting
concepts we now give the definition of an AGM in\defref{def-graph-agm}.
\begin{mydef}\label{def-graph-agm}
  An \emph{Abstract Graph Machine}(AGM) is a 6-tuple (G, \WorkItemSet, Q, \PF, $<_{\wis}$, S), where
  \begin{enumerate}
  \item G = (V, E) is the input graph,
  \item $\textbf{\WorkItemSet} \subseteq (V \times P_0 \times P_1 \dots \times P_n)$
where each $P_i$ represents a state value or an ordering attribute,
  \item Q - Set of states represented as mappings,
  \item $\PF : \WorkItemSet \longrightarrow \powerset{\WorkItemSet}$ is the processing function,
  \item $<_{\wis}$ - Strict weak ordering relation defined on \WorkItems
  \item S ($\subseteq \WorkItemSet$) - Initial \WorkItem set.
  \end{enumerate}
\end{mydef}

In the following we give the semantics of an AGM.
An AGM starts execution with the \emph{initial \WorkItem set}.
The initial \WorkItem set is ordered according to the strict weak
ordering relation. Next, the \WorkItems within the smallest equivalence class
is fed to the processing function.
If the processing function generates new \WorkItems,
then they are separated into equivalence classes based on the strict
weak ordering relation.
The \WorkItems within a single equivalence class can execute the
processing function
in parallel. However, \WorkItems in two different equivalence
classes must be ordered according to the induced relation (i.e. $<_{\WIS}$).
When executing \WorkItems in an equivalence class, it may generate
new \WorkItems for the same equivalence class or to an equivalence
class greater (as per $<_{\WIS}$)
than currently processing equivalence class.
The AGM executes \WorkItems in the next equivalence class,
once it finished executing all the \WorkItems in the
current smallest equivalence class. An AGM terminates when it
executes all the \WorkItems in all the equivalence classes.

The AGM is used to model graph algorithms related to
graph applications such as SSSP, BFS,
PageRank and Connected Components. Processing functions
and orderings used in those algorithms is discussed detail
in \citep{2016agm}.




\subsection{SSSP Algorithms in AGM}
\label{usecases}
In this subsection,
we build AGM model for \algoref{SSSP_relax}.
We also show that adding different
orderings to the modeled algorithm reveals behaviours
of existing distributed SSSP algorithms.

The SSSP algorithms discussed in this paper
can be formulated using a single statement.
For brevity we use following notation
to represent the processing function.

\begin{mynot}\label{notation}
  $\PF : \WorkItemSet \longrightarrow \powerset{\WorkItemSet}$
  \[ \PF(w) =  \begin{cases}
    \{ w_k | w_k \in <\Constructor(w)>,
    \\ \;\;\;<\StateUpdate(w)>,
    \\ \;\;\;<\Condition(w)>\}
  \end{cases}
  \]
\end{mynot}

In notation~\ref{notation}, $w_k$ is the new \WorkItem
generated from \Constructor. As discussed previously
\StateUpdate and \Condition represents state update
and condition.


To build the AGM for self-stabilizing~\algoref{SSSP_relax},
we need to define each tuple element in\defref{def-graph-agm}
(i.e. (G, \WorkItemSet, Q, \PF, $<_{\wis}$, S)).
As the input graph we use $G=(V, E, weight)$,
where \emph{weight} is a read-only property map
that has weights associated to edges.
As explained in\secref{ssandsssp} the set \WorkItemSet is defined
based on the vertex state. For\algoref{SSSP_relax}, the state we are
interested in is the distance from the source vertex. Therefore, we define,
$\WorkItemSet^{sssp}  \subseteq (V \times Distance)$,
where \emph{Distance} $\subseteq \mathbb{R}_+^*$.
The only state AGM needs is the distance from the source
vertex and it is represented as \emph{distance}.
Values for \emph{distance} state is assigned
when the processing function is executed with \WorkItems.

The AGM processing function is defined based on the rules
in\algoref{SSSP_relax} and given in\defref{sssp-pf}. Since Rule 1 is only applied to the source
vertex, we can move it to the initial \WorkItem set in the
AGM representation. Rule 2 is encoded into the processing function
in the format defined in\defref{pf}.
The definition of the processing function uses a helper
function called \emph{neighbors}
(Declared as $neighbors : V \longrightarrow \powerset{V}$), which operates on graph vertices.

\begin{mydef}\label{sssp-pf}
  $\PF^{sssp}:\WorkItemSet^{sssp} \rightarrow \powerset{\WorkItemSet^{sssp}}$
  \[ \PF^{SSSP}(w) =  \begin{cases}
    \{ w_k | w_k \in < w_k[0] \in \mli{neighbors}(w[0]) \; \mli{and}
    \\ \;\;\; w_k[1] = w[1] + \mli{weight}(w[0], w_k[0]) >,
    \\ \;\;\;< distance(w[0]) \longleftarrow w[1] >,
    \\ \;\;\;< \mli{if} \; w[1] < \mli{distance}(w[0]) >
    \\ \;\;\;\; \} \\
  \end{cases}
  \]
\end{mydef}

The next required
parameter definition for the AGM model
is the strict weak ordering relation ($<_{\wis}$).
If \WorkItems are not ordered in any form, then
we will have a single large equivalence class of generated
\WorkItems. We can have a single large equivalence class by
defining strict weak ordering relation as in\defref{chao-swo}.

\begin{mydef}\label{chao-swo}
  $<_{\mli{chaotic}}$ is a binary relation defined on $\WorkItemSet^{sssp}$ where,
  $w_1 <_{\mli{chaotic}} w_2 \; False \; \forall w_1, w_2 \in \WorkItemSet^{sssp}$
\end{mydef}

Basically, the binary relation $<_{\mli{chaotic}}$ does not divide
\WorkItems into any comparable equivalence classes.
Using the strict weak ordering defined in\defref{chao-swo}, we present the AGM model
for\algoref{SSSP_relax}, in\propref{chao-agm}.

\begin{myproposition}\label{chao-agm}
 \begin{enumerate}
 \item $G = (V, E, weight)$ is the input graph,
 \item \WorkItemSet = $\WorkItemSet^{sssp}$,
 \item Q = \{distance\} is the state mappings,
 \item \PF = $\PF^{SSSP}$,
 \item Strict weak ordering relation $<_{\wis}$ = $<_{\mli{chaotic}}$,
 \item S = \{\textless$v_s$, 0\textgreater\} where $v_s \in V$ and $v_s$ is the source vertex.
 \end{enumerate}
\end{myproposition}

Since the AGM presented in\propref{chao-agm},
does not perform any ordering on \WorkItems,
we call it \emph{Chaotic AGM}.
However, the ordering on \WorkItems can be improved by defining
strict weak ordering relations that generate smaller comparable
equivalence classes. There are numerous ways to define strict
weak orderings so that they generate smaller equivalence
classes. Some of those strict weak orderings yield us,
existing distributed SSSP algorithms such as \emph{Dijkstra's SSSP}\citep{dijkstra1959note} algorithm,
\emph{\Dstepping SSSP}\citep{meyer2003delta} algorithm and \emph{KLA SSSP}
algorithm\citep{harshvardhan_kla:_2014}.
All those algorithms share almost the same processing function
but uses different orderings on \WorkItems.

\subsubsection{Dijkstra's Algorithm}

Dijkstra's SSSP algorithm is the work efficient SSSP algorithm.
Dijkstra's algorithm globally orders vertices by their associated
distances and the shortest distance vertices are processed first.
In the following, we define the ordering relation
for Dijkstra's algorithm and
we instantiate Dijkstra's algorithm using Abstract Graph Machine.
\RO{The Dijkstra's algorithm description}

\begin{mydef}\label{dj-swo}
  $<_{\mli{dj}}$ is a binary relation defined on $\WorkItemSet^{sssp}$ as follows;
  Let $w_1, w_2 \in \WorkItemSet^{sssp}$, then;
  $w_1 <_{\mli{dj}} w_2$ iff $w_1[1] < w_2[1]$
\end{mydef}

It can be proved that $<_{dj}$ is a strict weak ordering that
satisfy constraints listed under Definition~\ref{dj-swo} (proof is omitted to save space).
AGM instantiation for Dijkstra's algorithm is given in~\propref{dijkstra-agm}.

\begin{myproposition}\label{dijkstra-agm}
  Dijkstra's Algorithm is an instance of AGM where;
 \begin{enumerate}
 \item $G = (V, E, weight)$ is the input graph,
 \item \WorkItemSet = $\WorkItemSet^{sssp}$,
 \item Q = \{distance\} is the state mappings,
 \item \PF = $\PF^{SSSP}$,
 \item Strict weak ordering relation $<_{\wis}$ = $<_{\mli{dj}}$,
 \item S = \{\textless$v_s$, 0\textgreater\} where $v_s \in V$ and $v_s$ is the source vertex.
 \end{enumerate}
\end{myproposition}

\subsubsection{\DStepping Algorithm}

\DStepping~\cite{meyer2003delta} SSSP algorithm
arranges vertex-distance pairs
into distance ranges (\emph{buckets}) of size $\Delta (\in \mathbb{N})$
and execute buckets in order.  Within a bucket, vertex-distance pairs
are not ordered,
and can be executed in any order.
Processing a bucket may produce extra work for
the same bucket or for a successive buckets.
The strict weak ordering relation for \Dstepping algorithm
is given in Definition~\ref{delta-swo}.
\RO{The delta stepping algorithm}

\begin{mydef}\label{delta-swo}
  $<_\Delta$ is a binary relation defined on $\WorkItemSet^{sssp}$ as follows;
  Let $w_1, w_2 \in \WorkItemSet^{sssp}$, then; \newline
  $w_1 <_{\Delta} w_2$ iff $\lfloor w_1[1]/\Delta \rfloor < \lfloor  w_2[1]/\Delta \rfloor$
\end{mydef}

Instantiation of $\Delta$-Stepping algorithm in the
AGM is same as in Proposition~\ref{dijkstra-agm},
except the strict weak ordering relation is $<_{\Delta}$ (= $<_{\wis}$).


\subsubsection{KLA SSSP Algorithm}

The $k$-level asynchronous (KLA) paradigm\citep{harshvardhan_kla:_2014}
bridges \emph{level-synchronous} and
asynchronous paradigms for processing graphs.
In level-synchronous approach (E.g:- In level-synchronous BFS),
vertices are processed level by level. KLA processes vertices
up-to $k$ levels asynchronously and then moves to next
$k$ levels.
\RO{What is KLA SSSP algorithm}

In the following we model KLA SSSP with the AGM. The KLA
approach order \WorkItems by the \emph{level}
of the resulting tree. Therefore, we need an additional
ordering attribute in the \WorkItemSet definition.
The KLA \WorkItemSet is defined as $\WorkItemSet^{kla} \subseteq V \times \mli{Distance} \times \mli{Level}$
where $Level \subseteq \mathbb{N}$.
\RO{\WorkItemSet for KLA algorithm}

Further, the processing function also needs to be altered to
populate value for the new ordering attribute.
The processing function for KLA SSSP is defined in Definition~\ref{kla-sssp-pf}.
The processing function generates \WorkItems with an updated level, while
changing distance state appropriately.
\RO {The processing function for KLA SSSP}

\begin{mydef}\label{kla-sssp-pf}
  $\PF^{kla} : \WorkItemSet^{kla} \longrightarrow \powerset{\WorkItemSet^{kla}}$
  \[ \PF^{kla}(w) =  \begin{cases}
    \{ w_k | w_k \in < w_k[0] \in neighbors(w[0]) \; and
    \\ \;\; w_k[1] = w[1] + weight(w[0], w_k[0])
    \\ \;\;\;\;\; and \; w_k[2] = w[2] + 1 >,
    \\ \;\;\;< distance(w[0]) \longleftarrow w[1] >,
    \\ \;\;\;\;\; < if \; w[0] < distance(w[0]) > \}
  \end{cases}
  \]
\end{mydef}

KLA SSSP order \WorkItems based on the $k$ level.
The \WorkItems within two consecutive $k$ levels can be executed in
parallel and \WorkItems that are not in two consecutive $k$ levels
must be ordered. The strict weak ordering relation for KLA
SSSP is given in Definition~\ref{kla-sssp-rel}.
\RO{Strict weak ordering for KLA SSSP}

\begin{mydef}\label{kla-sssp-rel}
  $<_{kla}$ is a binary relation defined on \newline
  $\WorkItemSet^{kla}$ as follows: \\
  Let $w_1, w_2 \in \WorkItemSet^{kla}$, then \newline
  $w_1 <_{kla} w_2$ iff $\lfloor w_1[2]/k \rfloor < \lfloor w_2[2]/k \rfloor$
\end{mydef}

The AGM formulation for KLA SSSP algorithm is given
in Proposition~\ref{kla-agm}.

\begin{myproposition}\label{kla-agm}
  KLA SSSP Algorithm is an instance of AGM where;
 \begin{enumerate}
  \item $G = (V, E, weight)$ is the input graph
  \item \WorkItemSet = $\WorkItemSet^t$
  \item Q = \{distance\} are the state mappings,
  \item \PF = $\PF^{kla}$
  \item Strict weak ordering relation $<_{\wis}$ = $<_{kla}$
  \item S = \{\textless$v_s$, 0, 0\textgreater\} where $v_s \in V$ and $v_s$
    is the source vertex and level starts with 0.
 \end{enumerate}
\end{myproposition}

\begin{figure}
         \begin{subfigure}[b]{0.20\textwidth}
                 \centering
                 \includegraphics[width=0.9\textwidth, height=1cm]{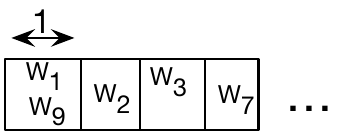}
                 \caption{Dijkstra's SSSP}
                 \label{fig2:djsssp}
         \end{subfigure}
         \begin{subfigure}[b]{0.20\textwidth}
                 \centering
                 \includegraphics[width=0.985\textwidth]{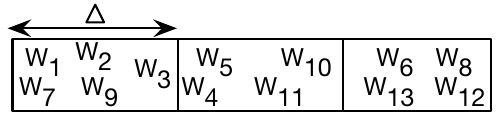}
                 \caption{\DStepping SSSP}
                 \label{fig2:deltasssp}
         \end{subfigure}%

     \begin{subfigure}[b]{0.20\textwidth}
             \centering
             \includegraphics[width=0.98\textwidth]{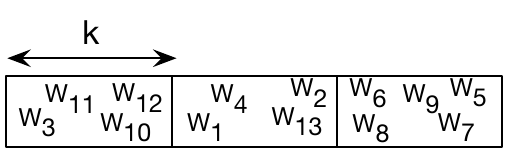}
             \caption{$k$-Level}
             \label{fig2:klassssp}
     \end{subfigure}
     \begin{subfigure}[b]{0.20\textwidth}
             \centering
             \includegraphics[width=0.985\textwidth]{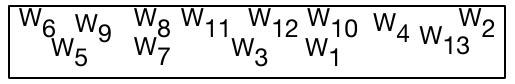}
             \caption{Chaotic (No ordering)}
             \label{fig2:chaoticsssp}
      \end{subfigure}
\caption{Equivalence classes generated by different SSSP algorithms}
\label{fig:equiclasses}
\end{figure}

Each AGM discussed above divides \WorkItemSet into
equivalence classes differently (Figure~\ref{fig:equiclasses})
The Dijkstra's AGM (Figure~\ref{fig2:djsssp}) generates an
equivalence class per each different distance value. The \WorkItems
that have the same distance belong to the same equivalence class while \WorkItems
of different distances go into different equivalence classes.
The \Dstepping (Figure~\ref{fig2:deltasssp}) AGM
also performs ordering based on the distance but
equivalence classes are generated based on a $\Delta$ value and
in general have more elements compared to Dijkstra's AGM. All
the \WorkItems within a single equivalence class are guaranteed to
have distances between $\Delta * i$ and $\Delta * (i+1)$ for
some $i \in \mathbb{N}$.
Similar to \Dstepping algorithms, KLA, too,
generates larger equivalence classes but uses level as
the ordering attribute.
Figure~\ref{fig2:klassssp}, shows how KLA algorithm
arranges equivalence classes. As shown in Figure~\ref{fig2:chaoticsssp},
the Chaotic version has a single large equivalence class containing all
the \WorkItems.

Each of the above algorithms is different because of the way
they generate equivalence classes and how those equivalence
classes are ordered. Otherwise, they tend to
share the same processing function that implements Rule 1
in\algoref{SSSP_relax}. Further, if two SSSP algorithms
share the same ordering attribute, then they share
the same processing function in AGM model. For example, both Dijkstra's
AGM and \Dstepping share the same processing function.
If two algorithms use different ordering attributes, then
processing functions differ only to update values of
different ordering attributes.
\AJ{Result 1}

\section{Extended-AGM}
\label{sec:extended-agm}

Ordering in terms of distance reduces the amount of redundant
work in SSSP algorithms. Out of the algorithms discussed in the
previous section, Dijkstra's algorithm performs the best ordering,
so that it does the minimum amount of redundant work. However, the overhead of ordering
in Dijkstra's algorithm
is significant in a parallel distributed run-time due to the frequent
synchronization. In other words, when the AGM instance generates
more equivalence classes, the global synchronization overhead increases.
The other
algorithms discussed above reduce overhead of ordering by
chunking \WorkItems into larger equivalence classes. This reduces
the number of equivalence classes generated.
The \emph{Extended-AGM} adds ordering to chunks
but to reduce the overhead of ordering at AGM level, it applies
ordering at different lower \emph{spatial levels}.

\AJ{MZ, MB : Do we have a better way to explain what a spatial level is ?}
A spatial level defines the amount of memory accessible to
order \WorkItems. The highest spatial level is the accumulation
of all the memory of the participating nodes (also called \emph{global memory}).
The next spatial level is the memory available at a node.
The memory in a single node can be further subdivided into logical regions such
as \emph{numa} domains.
Each numa domain may be shared by several threads.
The lowest spatial level is the thread local memory.
Such a spatial division is highly architecture-dependent and
hierarchical.

\begin{wrapfigure}{L}{.30\linewidth}
\centering
\includegraphics[width=\linewidth]{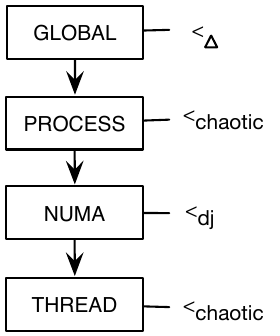}
\caption{EAGM Spatial Hierarchy}
\vspace{-1ex}
\label{fig:spatial-hierarchy}
\end{wrapfigure}

The EAGM depicts a spatial hierarchy as a tree (\figref{fig:spatial-hierarchy}).
Every node in the spatial hierarchy has an
ordering attached. The ordering
attached to the root represents the ordering
defined by the relevant AGM.
The example given in\figref{fig:spatial-hierarchy},
is an EAGM hierarchy derived from the \Dstepping AGM.
As shown in\figref{fig:spatial-hierarchy}
the $\Delta$ ordering (\defref{delta-swo})
is attached to the root node.
The orderings attached to the lower spatial
levels are performed on \WorkItems available to
the memory in the relevant spatial level.
Therefore, the orderings attached to lower spatial levels
are more \emph{relaxed} than the ordering attached to the root.
By default the EAGM spatial hierarchy assumes
Chaotic (i.e. no ordering) ordering but
specific orderings can be enforced. The example
in \figref{fig:spatial-hierarchy}
enforces strict weak ordering $<_{dj}$
at numa level and uses Chaotic orderings
at other levels.

An \emph{Extended AGM} is an AGM but
instead of consisting a strict weak ordering relation
an EAGM has a spatial hierarchy with annotated
orderings. An EAGM extends an AGM if and only if the EAGM generates
same equivalence classes in the AGM at the root
level of the spatial hierarchy. Therefore, an AGM can have multiple
EAGMs, where each EAGM has the same ordering as the AGM
at the root but different orderings at lower
spatial levels. Each EAGM represents a variation of the algorithm
modeled in the relevant AGM. If an AGM
generates equivalence classes with many \WorkItems, then
the EAGM has provision to perform fine-grain ordering at different
spatial levels. However, if the AGM ordering is such that it
generates equivalence classes with few \WorkItems,
then the derived EAGM has less opportunity to perform ordering
at spatial levels.
For example, the \Dstepping AGM generates equivalence
classes with many \WorkItems (provided $\Delta$ is sufficiently large).
Variations of the \Dstepping AGM can be generated by applying
ordering to \WorkItems at the process level, the numa level
or at the thread level. However, for Dijkstra's AGM the spatial orderings
may not improve the overall performance of the algorithm because
the equivalence classes generated by Dijkstra's AGM has fewer \WorkItems
on average.

\begin{figure}
  \centering
  \begin{subfigure}{\linewidth}
    \includegraphics[width=\linewidth]{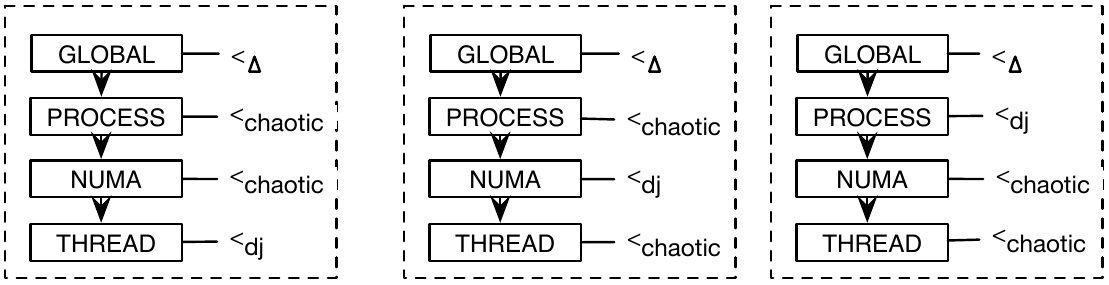}
    \caption{EAGMs of \Dstepping AGM}
    \label{cw_10}
  \end{subfigure}
  \begin{subfigure}{\linewidth}
    \includegraphics[width=\linewidth]{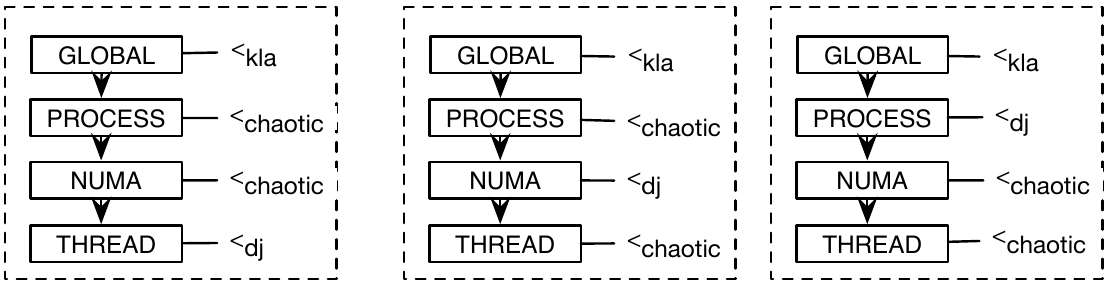}%
    \caption{EAGMs of KLA AGM}
    \label{cw_25}
  \end{subfigure}
  \begin{subfigure}{\linewidth}
    \includegraphics[width=\linewidth]{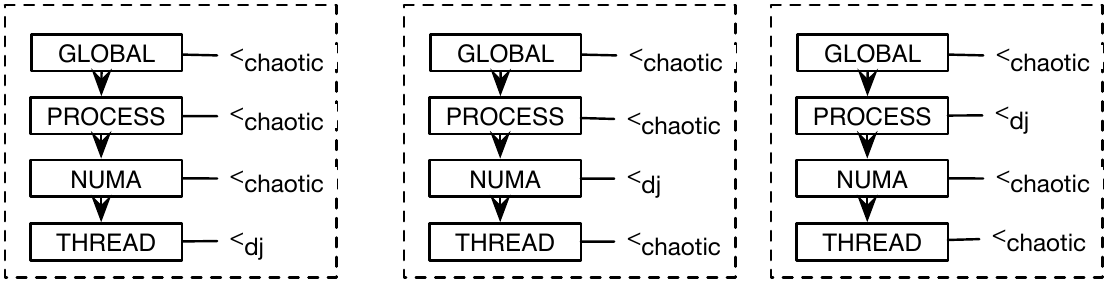}%
    \caption{EAGMs of Chaotic AGM}
    \label{cw_50}
  \end{subfigure}
  \caption{Thread ordered, NUMA ordered and Process ordered EAGMs for \Dstepping, KLA and Chaotic AGMs.}
  \label{fig:nineeagsm}
\end{figure}

Out of the algorithms discussed in the previous
section, the fine-grained spatial
ordering is effective to AGMs defined for \Dstepping,
KLA and Chaotic. By considering the spatial hierarchy
used in Figure~\ref{fig:spatial-hierarchy}, we apply
Dijkstra's strict weak ordering relation (\defref{dj-swo}) to
spatial hierarchy levels \emph{PROCESS},
\emph{NUMA} and \emph{THREAD} to
derive EAGMs (Figure~\ref{fig:nineeagsm}).
Each EAGM generates a variation of the main algorithm
defined by its corresponding AGM.

Figure~\ref{cw_10}, shows EAGM variations derived for
\Dstepping algorithm. Figure~\ref{cw_10}-(i),
applies $<_{dj}$ ordering to THREAD level and
Figure~\ref{cw_10}-(ii) applies $<_{dj}$
ordering to NUMA level and Figure~\ref{cw_10}-(iii)
applies ordering to PROCESS level. EAGMs for
KLA and Chaotic are derived in the same way.

For convenience, we will refer to the original AGM
implementation as \emph{buffer}, the variation
that does THREAD level ordering as \emph{threadq},
the variation that does PROCESS level ordering as
\emph{nodeq} and the variation that does NUMA
level ordering as \emph{numaq}.

\section{Implementation}
\label{sec:impl}

We implemented EAGMs shown in Figure~\ref{fig:nineeagsm}.
Each implementation generated a variation of main
algorithm. For implementation
we used a light weight active messaging
framework based on MPI.
To represent the local graph data,
we used \emph{compressed sparse row} format.
1D distribution is used to distribute
the graph. States and read only mappings are maintained
as property maps indexed by vertices or edges and property maps are also
distributed.
We used concurrent priority queues
(\emph{flat combining synchronous priority queue}\citep{hendler2010flat}) for
node level and numa level ordering.



\begin{figure*}
  \centering
\includegraphics{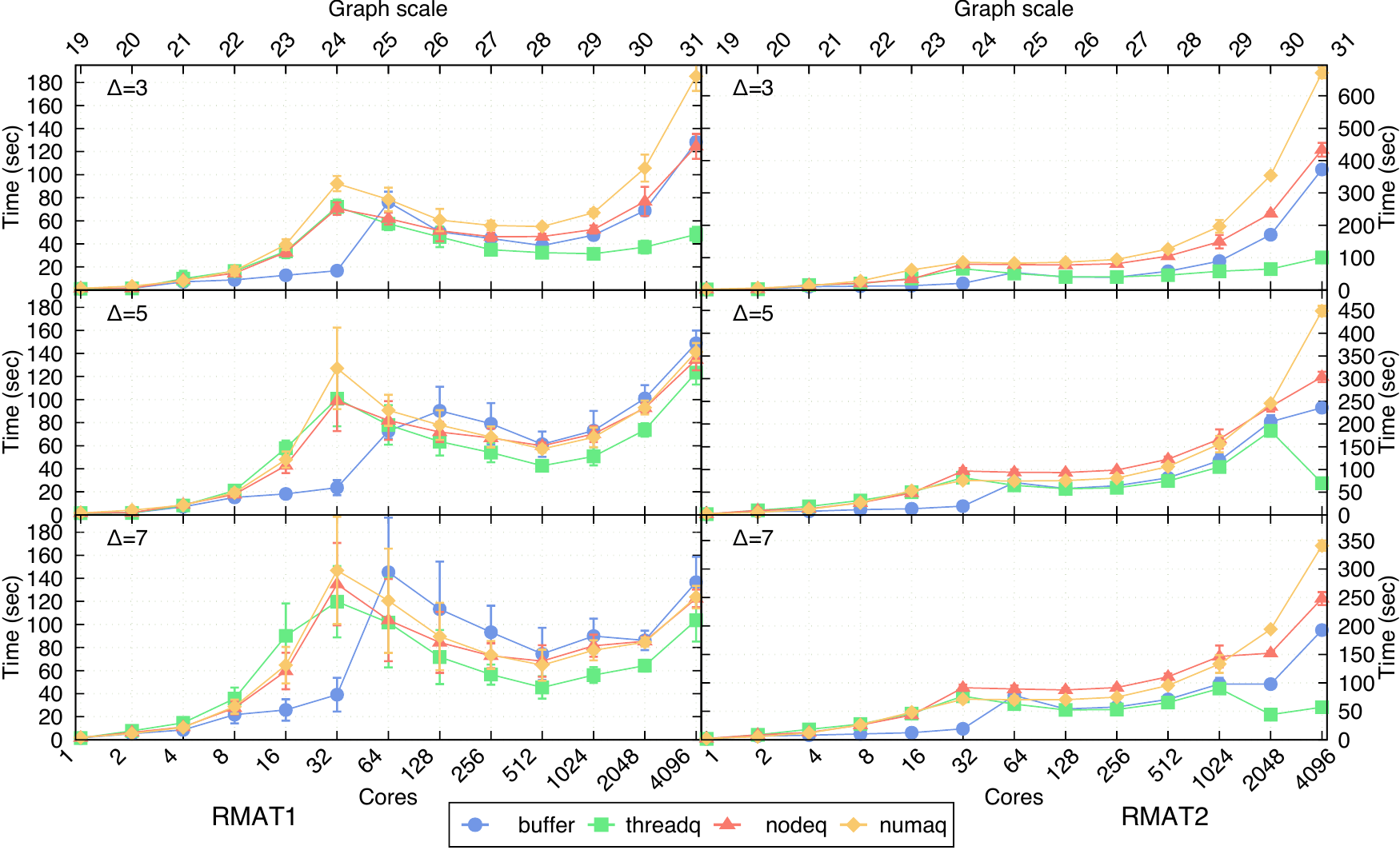}
\caption{Timing results of \Dstepping variations for RMAT1 and RMAT2 graphs, with $\Delta=3$, $\Delta=5,$ and $\Delta=7.$}
\label{plot:ds_results}
\end{figure*}

\section{Results}

\begin{figure*}
  \centering
  \includegraphics{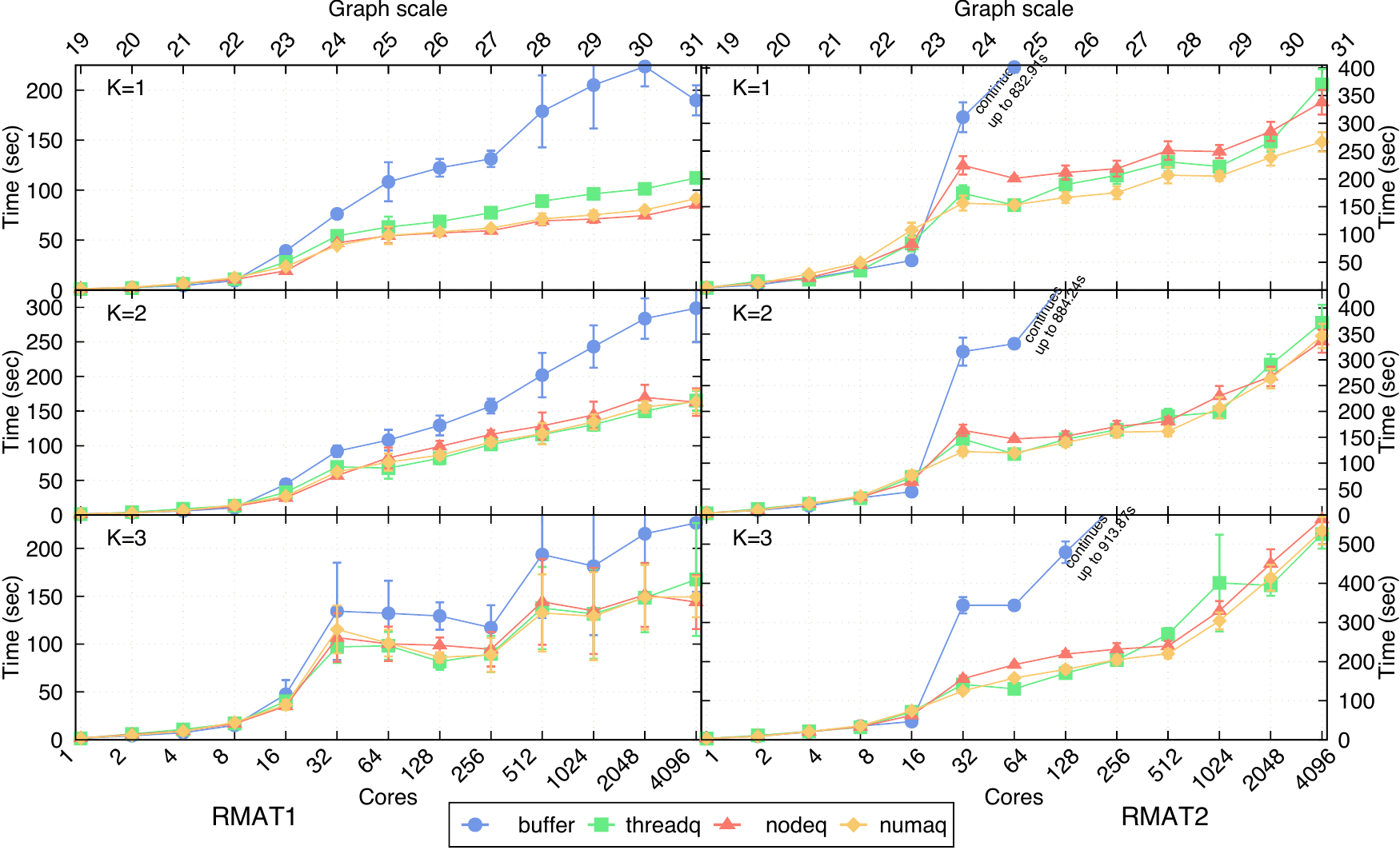}
  \caption{Timing results of KLA variations for RMAT1 and RMAT2 graphs, with $K=1$, $K=2,$ and $K=3.$}
  \label{plot:kla-results}
\end{figure*}

We experimented with the performance of EAGM variations on synthetic graphs and
on real-world graphs. For synthetic graphs we used:
\begin{itemize}
\item RMAT1: Graphs based on the current Graph500\citep{murphy2010introducing}
  BFS benchmark specification with \emph{R-MAT}\citep{chakrabarti2004r}
  parameters $A = 0.57$, $B = C = 0.19$ and $D = 0.05$ and random edge weights
  from $1$ to $100$.
\item RMAT2: Graphs generated based on the proposed Graph500\citep{graph500sssp}
  SSSP benchmark specification with R-MAT parameters $A = 0.50$, $B = C = 0.1$
  and $D = 0.3$ and random edge weights from $1$ to $255.$
\end{itemize}
For real world graphs, we use four graphs with varying parameter from the
SNAP\citep{snapnets} repository with random edge weights from $1$ to $100.$

All our experiments were carried out on a Cray XE6/XK7 supercomputer, with 32 AMD Opteron Abu Dhabi
CPUs 64 GB memory per node (4 NUMA domains), running Cray Linux Environment and the cray-mpich2 ver. 7.2.5 MPI implementation.  The run-time architecture we used, aligned with
the spatial hierarchy used in our EAGM implementations
(Figure~\ref{fig:nineeagsm}).


\subsection{Scaling Results}

\begin{figure*}
  \centering
  \includegraphics{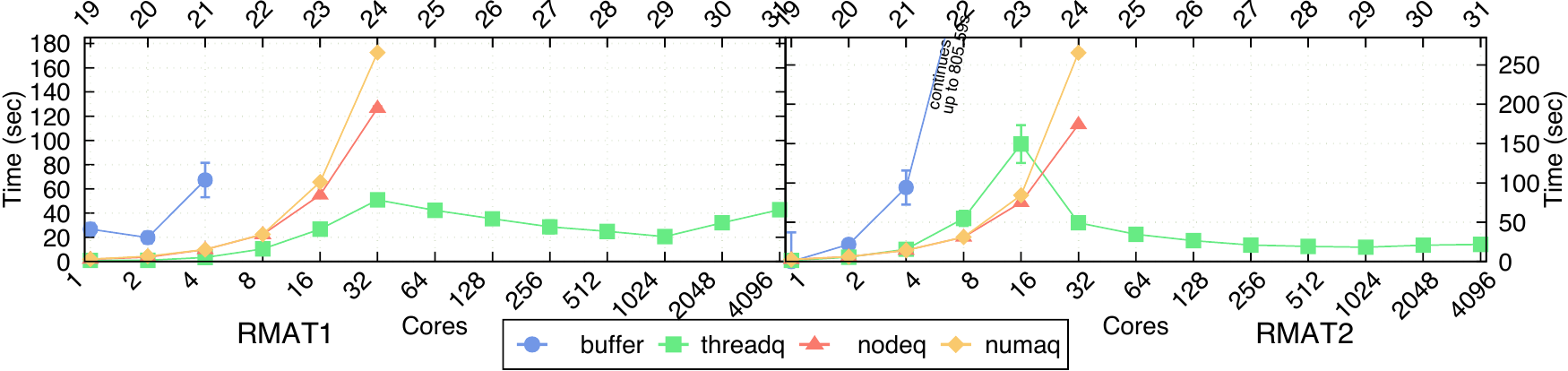}
\caption{Timing results of chaotic variations for RMAT1 and RMAT2 graphs.}
\label{plot:chaotic-results}
\end{figure*}

We ran weak scaling experiments on Graph500 graphs from scale 19 ($2^{19}$
vertices) to scale 31.  We compare each AGM algorithm (i.e. main SSSP algorithms)
to their EAGM variations (see
\cref{sec:extended-agm}).
The implementation of the main AGM algorithm
is represented in \textbf{buffer} for each algorithm ($\Delta$-Stepping,
KLA and Chaotic).
The thread-level ordering variation, node
level ordering variation and numa level ordering
variation implementations are denoted
using \textbf{threadq}, \textbf{nodeq}
and \textbf{numaq}, for each algorithm.



\subsubsection{\DStepping Variations}

In \cref{plot:ds_results}, we present weak scaling results for \Dstepping
variations on RMAT1 and RMAT2 synthetic graphs with $\Delta$ values 3, 5, and 7.
The original \Dstepping (buffer) algorithm performs the best in-node.  Since no communication is
involved, the additional ordering provided by the other implementations does not
provide a sufficient benefit for its overhead.

In general,  the threadq
variation is the fastest in the distributed setting.  The nodeq and the numaq variations perform better with increasing deltas, and they are not competitive with the buffer implementation.

In summary, while in-node performance is dominated
by the traditional \Dstepping algorithm (aka the buffer implementation) the distributed
execution shows significant improvement with threadq
variation.  Though, the numaq and nodeq variations should
provide better performance than the threadq variation by providing more ordering,
the overhead of the concurrent ordering reduces the performance of
numaq and nodeq.




\subsubsection{KLA Variations}

KLA variations show different performance
characteristics than \Dstepping. Weak scaling
results for KLA implementations on RMAT1 and RMAT2 graphs $K=1,2$ and $3$ are shown in \cref{plot:kla-results}.
For KLA, the nodeq and the numaq variations perform the best at scale, with $K=1.$  At greater $K$ values, the performance of threadq is comparable to nodeq and numaq, but in absolute terms, the performance at higher $K$ values is worse than at $K=1.$   As said in the previous section, the numaq and nodeq provide the best potential ordering by ordering the most items.  The overheads are kept at bay because at $K=1$ all the writes to the next level's queue occur before all the reads.  The flat combining queue\citep{hendler2010flat} that we use performs the best in this scenario.  For higher $K$ values, writes and reads get more mixed, and the advantage of numaq and nodeq becomes less pronounced (when $K$ is higher more \WorkItems go into queues and concurrent ordering overhead become significant).
In KLA, for both RMAT1 and RMAT2 inputs,
all EAGM variations (threadq, nodeq and numaq) perform better compared to original KLA algorithm (buffer).



\begin{table*}
\begin{center}
\footnotesize
\begin{tabular}[l]{ m{1.2cm} m{1.1cm} m{1.1cm} m{0.3cm} m{0.5cm} l l l l l l l l l}
Graph & $|V|$ & $|E|$ & $D$ & Cores & AGM & buffer & $\sigma_{buf}$ & threadq & $\sigma_{thread}$ & nodeq & $\sigma_{node}$ & numaq & $\sigma_{numa}$ \\
\hline
\multirow{2}{1.2cm}{SOC-Live\citep{leskovec2009community}} & 4847571 & 68993773 & 16 & 64 & $\Delta=3$ & 22.7 & 6.47 & 20.52 & 5.8 & 14.93 & 2.39 & 29.8 & 3.87 \\
& & & & & $KLA,\; K=1$ & 90.39 & 5.78 & 38.66 & 4.33 & 25.08 & 1.94 & 35.96 & 2.55 \\
& & & & & $Chaotic$ & 39.82 & 6.98 & 11.66 & 0.63 & 166.26 & 15.42 & 207.42 & 22.21 \\
\hline
\multirow{2}{1.2cm}{Wiki-Talk\citep{leskovec2010signed}} & 2394385 & 5021410 & 9 & 64 & $\Delta=3$ & 2.27 & 0.28 & 2.26 & 0.57 & 3.44 & 0.28 & 9.71 & 0.5 \\
& & & & & $KLA,\; K=1$ & 3.73 & 0.44 & 2.53 & 0.34 & 1.94 & 0.23 & 3.9 & 0.4 \\
& & & & & $Chaotic$ & 41.46 & 6.54 & 1.41 & 0.05 & 8.34 & 1.29 & 5.97 & 0.78 \\
\hline
\multirow{2}{1.2cm}{roadNet-CA\citep{leskovec2009community}} & 1965206 & 2766607 & 849 & 1024 & $\Delta=1200$ & 24.53 & 6.49 & 22.28 & 6.09 & 12.23 & 1.2 & 19.18 & 2.05 \\
& & & & & $KLA,\; K=10$ & 54.24 & 6.14 & 54.63 & 5.8 & 43.86 & 5.18 & 51.35 & 5.54 \\
& & & & & $Chaotic$ & 44.62 & 7.4 & 2.68 & 0.21 & 44.17 & 2.08 & 23.76 & 2.1 \\
\hline
\multirow{2}{1.2cm}{Orkut\citep{yang2015defining}} & 3072441 & 117185083 & 9 & 1024 & $\Delta=10$ & 3.72 & 0.43 & 3.12 & 0.4 & 4.29 & 0.36 & 6.77 & 0.3 \\
& & & & & $KLA,\; K=5$ & 20.94 & 6.33 & 71.19 & 23.5 & 15.87 & 1.56 & 18.28 & 2.12\\
& & & & & $Chaotic$ & 64.84 & 12.19 & 2.97 & 1.14 & 56.81 & 5.43 & 50.41 & 3.99 \\
\end{tabular}
\end{center}
\vspace{-1ex}
\caption{Mean timing results (in seconds) for buffer, threadq, nodeq and numaq when ran against listed real graphs.}
\label{table:realtime}
\end{table*}

\subsubsection{Chaotic Variations}

\Cref{plot:chaotic-results} shows results for chaotic variations with RMAT1 and
RMAT2 input graphs.  The chaotic AGM has a single large equivalence class and
does not perform any form of ordering.  Due to work explosion we were unable to
run the chaotic algorithm except for small scales. For the same reason both
nodeq and numaq end up having larger queue sizes, hence the overhead of ordering
became significant as we increase the scale.  However, the thread level ordering
shows good performance, specially in distributed execution.  For RMAT2, threadq
achieves almost perfect weak scaling.  Furthermore, the threadq chaotic variation is faster than all other variations in terms of absolute performance, demonstrating how the structured (E)AGM approach may result in new, highly performant algorithms.

\subsection{Real World Graphs}

The real world graphs we used in our experiments are listed in
Table~\ref{table:realtime} along with their characteristics and relevant
results.  Most of the real world graphs are fairly small.  For our experiments,
we pick either 64 or 1024 cores, depending on the size of the graph.  SOC-Live
Journal\citep{leskovec2009community} and Wiki-Talk\citep{leskovec2010signed}
experiments are run on 64 cores, California Road
Network\citep{leskovec2009community}, and
Orkut\citep{yang2015defining} experiments are run on 1024 cores.

For Live Journal, nodeq showed better results out of the
\Dstepping variations and the $KLA$ variations. Similarly to
synthetic graph results, the chaotic variation show best results
for threadq. KLA nodeq and chaotic threadq showed the best results
on Wiki-Talk graph.
California Road Network has the highest diameter, with
the edge weights ranging from 0 to 100.  Both \Dstepping and KLA
show good results with the node level ordering
and Chaotic showed good results for threadq.
Orkut graph input shows minimum
timing values for \Dstepping in their threadq implementations.
In Chaotic
variations, threadq implementation continued to perform better for social
networking graphs.

\section{Related Work}
SSSP is a classic example of an
irregular application. Parallel graph
algorithms for SSSP being well-studied.
$\Delta$-Stepping~\cite{meyer2003delta}, KLA~\cite{harshvardhan_kla:_2014},
Bellman Ford~\cite{bellman1956routing}, Crauser's SSSP~\cite{crauser1998parallelization}
are popular algorithms that
address distributed memory parallel
SSSP problem.

A distributed SSSP algorithm connecting self-stabilizing
was studied in~\cite{zalewski_distributed_2014}.
The same algorithm is discussed for
different run-time characteristics in\citep{firozcomparison}.
Algorithm discussed
in~\cite{zalewski_distributed_2014} is an EAGM instantiation of Chaotic
AGM. Much of the work related to
self-stabilization is already discussed in\secref{sec:intro}.

AGM is an abstract model for designing distributed
memory parallel graph algorithms. Most
of the graph algorithms existing today are
designed based on PRAM architecture.
In PRAM we have a single shared memory
and individual processors reading/writing
from/to shared memory.

As discussed
in\secref{sec:intro}, PRAM algorithms suffer from
performance issues in distributed memory settings.
\textit{Bulk Synchronous Parallel} (BSP)~\cite{gerbessiotis1994direct} is
a model used for designing distributed
memory parallel algorithms. In BSP we have
super steps where in each super step we
do computation, communication and barrier synchronization.
Also, there are certain variations of BSP
where computation and communication are overlapped
to improve performance yet uses barriers.
BSP is more like an extended version of PRAM and
hence the graph algorithms designed for PRAM
can be implemented in distributed settings
using BSP. In addition, there are
models that consider network parameters
and considers communications (\eg LogP~\cite{culler1993logp}
and its descendants).


Regarding spatial orderings, Galois scheduler; OBIM~\cite{lenharth2011priority}
considers spatial features when processing an irregular application
but it is an implementation for shared memory systems
rather than an abstract model.


\section{Conclusion}

Using the AGM abstraction, we converted\algoref{SSSP_relax} to a form
suitable for distributed memory, parallel graph processing.
We showed that existing distributed graph algorithms; Dijkstra's SSSP,
\Dstepping SSSP and KLA SSSP are variations of the converted algorithm.
Those algorithms basically implement the Rule 2 of\algoref{SSSP_relax}
with different ordering based on distance state or based on
a different ordering attribute such as level.
We also showed, proposed EAGM model can generate more
fine-grained orderings at less synchronized spatial levels.
Results of our experiments showed that some of the generated
algorithms perform better compared to standard distributed memory,
parallel SSSP algorithms under different graph inputs.

\section*{Acknowledgment}
This work is supported by the National Science Foundation under Grant No. 1319520.
Further, this research was supported in part by Lilly Endowment, Inc., 
through its support for the Indiana University Pervasive Technology Institute, 
and in part by the Indiana METACyt Initiative. 
The Indiana METACyt Initiative at IU was also supported in part by Lilly Endowment, Inc. 

\bibliographystyle{IEEEtran}
\bibliography{IEEEabrv,paper}





%

\end{document}